# The ECMWF Ensemble Prediction System: Looking Back (more than) 25 Years and Projecting Forward 25 Years

By

T.N.Palmer[1,2]
[1]Department of Physics, University of Oxford
[2]European Centre for Medium-Range Weather Forecasts

## Abstract

This paper has been written to mark 25 years of operational medium-range ensemble forecasting. The origins of the ECMWF Ensemble Prediction System are outlined, including the development of the precursor real-time Met Office monthly ensemble forecast system. In particular, the reasons for the development of singular vectors and stochastic physics – particular features of the ECMWF Ensemble Prediction System - are discussed. The author speculates about the development and use of ensemble prediction in the next 25 years.

1. ## Introduction

The notion that the wintertime atmosphere has a "limit of deterministic predictability" of about two weeks arose from numerical experiments with general circulation models in the 1960s (e.g. Smagorinsky 1963, Leith 1965, Mintz, 1965). In these experiments the growth of small perturbations was studied for various synoptic flow types. Whilst the earlier work of Lorenz (1963) indicated that deterministic prediction would eventually be limited by the chaotic nature of the atmosphere, these experiments suggested that it should be possible to make useful deterministic forecasts at least 10 days ahead. This provided the scientific basis for the establishment of the European Centre for Medium-Range Weather Forecasts (ECMWF) in the 1970s, and for medium-range prediction more generally.

However, over the last 25 years, weather forecasting within this deterministic limit has undergone a radical change: from a deterministic procedure where at each forecast initial time a single prediction is made from a best-guess set of initial conditions using a best-guess deterministic computational representation of the underlying equations of motion, to a probabilistic one where an ensemble of predictions is made from a sample of initial conditions using stochastic computational representations of the underlying equations of motion. Thus, paradoxically for some, forecasting within this so-called deterministic limit of predictability has become inherently probabilistic. This paper focusses on the development of the ECMWF medium-range Ensemble Prediction System (EPS) (Palmer et al, 1992; Molteni et al, 1996).

Why were such developments necessary? As discussed in Section 2, in a chaotic system, the butterfly effect - here simply meant as the rapid growth of initial error - is itself flow dependent and hence will be manifest from time to time

*within* the limit of deterministic predictability. If we aspire to creating reliable forecast systems (i.e. systems which a user can rely on for making decisions), methods are needed to determine when the butterfly effect will compromise the accuracy of the forecast - even within the deterministic limit. A precursor to the ECMWF EPS is discussed in Section 3 – the Met Office real-time monthly ensemble forecast system. A key point here is that probabilistic prediction on the monthly timescale fitted naturally into the probabilistic statistical-empirical techniques that were being used to predict on the monthly timescale. By contrast, as discussed, ensemble techniques did not fit naturally into the prevailing methodologies for medium-range prediction. The specific ideas and research that led to the use of singular vectors and stochastic parametrisation in the EPS is discussed in Section 4. The explosion of development of ensemble prediction on all timescales is described and some of the challenges that remain are discussed in Section 5. In Section 6, an attempt in made to project forward to speculate how ensemble forecasting will be used routinely 25 years from now.

An important point must be recognised from the outset: because the EPS has been a development of the deterministic approach at ECMWF, it has relied on software developed primarily for deterministic prediction. In principle, it needn't have been like this. Conceivably, a probabilistic approach to numerical weather prediction could have followed the route of solving the Liouville or Fokker-Planck equations (Epstein, 1969). In this case, one would have had no option but to start again, pretty much from scratch. However, as discussed in Ehrendorfer (2006), such an approach would have been completely impracticable: for example, although the Liouville equation is linear in probability density, it is effectively infinite dimensional and cannot be reliably approximated by low-order truncations. Hence, the fact that the ECMWF EPS is world leading does draw on the fact that the deterministic ECMWF model and data assimilation systems are themselves world leading. In this sense, the development of the ECMWF EPS reflects the work of the whole of ECMWF, and not just the small group who were specifically focussed on the EPS (see Acknowledgements below). It is likely that model and data assimilation development will, in the next decade, be done entirely within an inherently probabilistic framework (recognising the "Primacy of Doubt"; Palmer, 2017). However, for now, the quality of ensemble systems does depend on the quality of the corresponding deterministic systems.

Since much of the discussion in this paper reflects the author's recollections and opinions, some of the text below will be written using the first-person singular.

## 2. Why Bother?

It is worth recalling some of the lively discussions and debates that were had, not only during the development phase of the EPS, but even more so when the EPS began to be run operationally and when it became apparent that the computational cost of the ensemble was not going to be trivial and would therefore be competing with resources for developing the traditional deterministic forecast. A key argument was that because medium-range forecasting was within the deterministic limit, we should focus the lion's share of available resources on making the deterministic forecasts as skilful as possible. In this way, it was argued, occasional forecast busts would be eventually eradicated. Why jeopardise ECMWF's status as the world-leading deterministic forecast centre, so the argument went, by diverting valuable human and computing resources to a probabilistic system that some felt would be unusable in practice?

The problem with this argument is that the notion of the two-week deterministic limit is itself a statistical one: all it means is that, on average, one can make useful deterministic predictions within this limit. However, because the climate system is nonlinear, the butterfly effect will, from time to time, destroy the accuracy of particular deterministic predictions on timescales much shorter than the deterministic limit. This concept is illustrated in Fig 1 which shows some finite-time ensemble integrations of the iconic Lorenz (1963) model. Lorenz's motivation for developing this model was that it provided a counter-example to the claim that deterministic long-range forecasting using empirical methods would be possible once we had enough independent observations to perform some sort of analogue prediction. However, Lorenz's model can also be used to show the Achilles Heel of deterministic prediction within the deterministic limit – quite a different application to what Lorenz originally had in mind. In Fig 1 is shown the growth of errors within the limit where deterministic predictions show, on average, some level of skill. The key point here is that since in the equations of motion

$$\dot{X} = F[X]$$

the functional $F$ is nonlinear, then necessarily the Jacobian $DF/DX$ in the linearised equation

$$\frac{d\delta X}{dt} = \frac{DF}{DX} \delta X$$

must necessarily depend (at least linearly) on $X$. This means that in a nonlinear system, the growth of small errors will vary with initial state.

Fig 1c shows an example of a situation where error growth is explosive. An example illustrating this situation in practice is the famous October 1987 storm over Southern England. Its misforecast led BBC Anchor Man Michael Buerk, the morning after the storm, to comment to the on-duty weather forecaster Ian McCaskill (https://www.youtube.com/watch?v=h0b92WZLlQw):
 "Well Ian, you chaps were a fat lot of good last night….if you can't forecast the worst storms for several centuries three hours before they happen, what are you doing!".

This point illustrates a crucial concept. Can we meaningfully say that deterministic forecasts are, on average, useful, when a small number of them turn out to be badly wrong? When a few forecasts go wrong, users begin asking themselves whether they can trust the forecast system *at all*. That is to say, a few poor forecasts compromise the reliability of the forecast system (for decision making) as a whole. (This concept is incorporated in the use of entropic measures of probabilistic skill, where forecast systems are severely penalised when the verification lies outside the forecast range, even on a single occasion.)

Of course, the converse (`crying wolf' situation) – where forecasters warn of some severe event (e.g. a category five hurricane or unprecedented snow amounts) and nothing exceptional actually occurs - can also lead users to question the value of forecast systems.

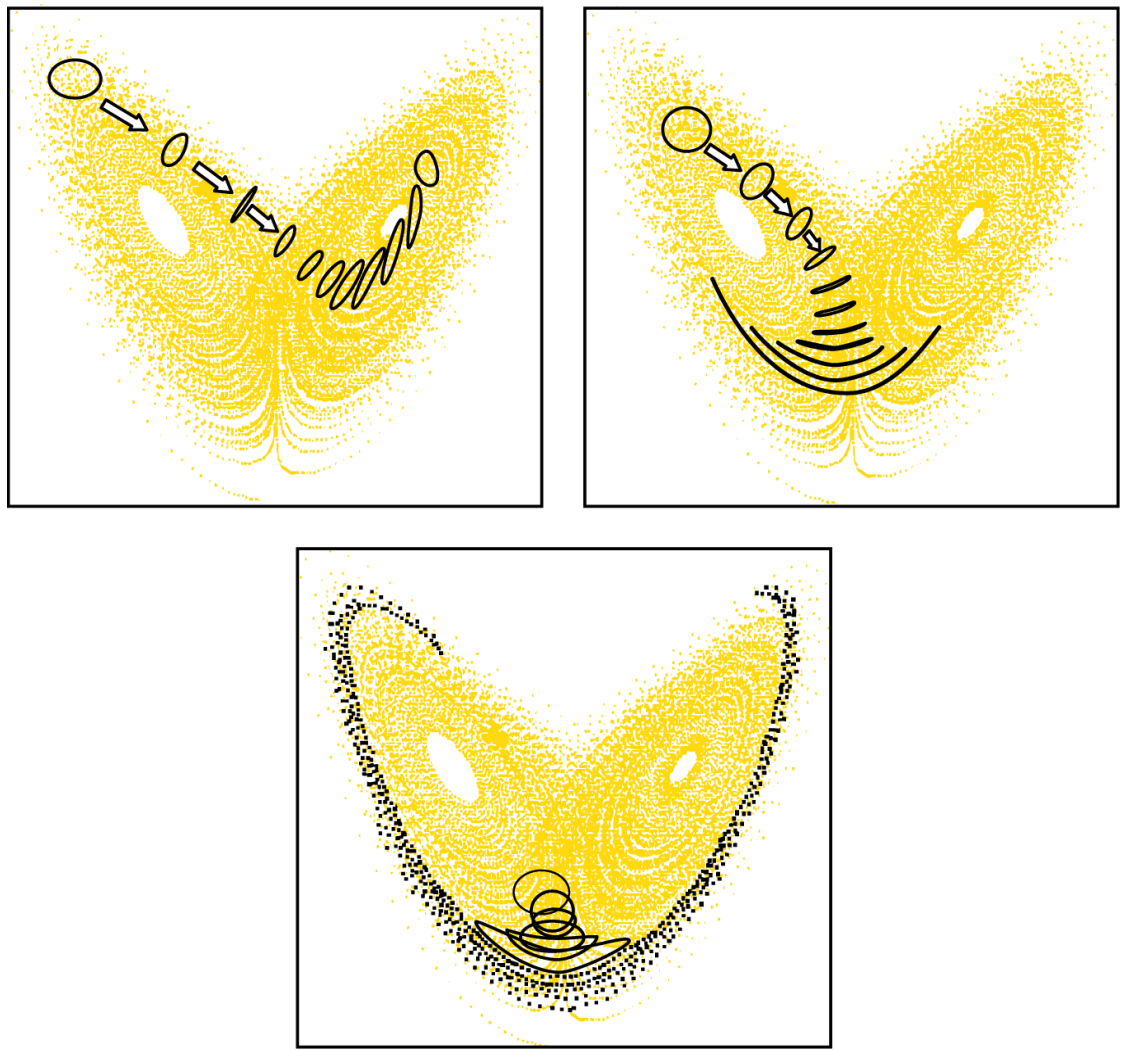

*Figure 1. Although Lorenz himself developed his iconic 1963 model to show the impossibility of deterministic long-range forecasting, it can also be used to illustrate the existence of an intermittent butterfly effect within the limit of deterministic predictability. For users to have confidence in forecasts within this deterministic limit, it is necessary to flag situations where this intermittent butterfly effect is active, and to predict plausible alternative weather scenarios probabilistically – more generally to predict flow-dependent uncertainty. This means abandoning the deterministic paradigm that has guided the development of numerical weather prediction over many years.*

Fig 1c, and the corresponding meteorological events, illustrate the fact that the so-called butterfly effect – here meaning sensitive dependence on initial conditions – can certainly occur intermittently within the limit of deterministic predictability. Improving the model, e.g. by increasing resolution, is not going to help solve this problem, indeed it may exacerbate it as the model resolves better the instability which causes the problem in the first place. Essentially Fig 1 illustrates the fact that any deterministic weather forecast system, no matter how good the forecast model, will have an Achilles Heel, even when forecasting within the deterministic limit.

There is another reason why forecast failures matter: when we say something will happen and it doesn't, our science is undermined. This bothers me a lot. By virtue of the fact that we make predictions every day, the science of meteorology satisfies Popper's criterion of falsifiability *par excellence.* However, if these predictions turn out to be false from time to time, then at best it means that there are aspects of our science we do not understand and at worst it provides ammunition to those that would see meteorology as empirical, inexact and unscientific. Indeed, manifest forecast failures do give ammunition to those who claim that predictions of climate change cannot be relied on because the forecasts we make on timescales where verification data exists are, on occasion, wrong.

I myself *do* think that meteorology is an exact science – but its exactitude lies in describing the evolution of probability distributions, not deterministic states. For those who think this is a "cop-out", we wouldn't say to a high-energy physicist at that her subject was not an exact science because the fundamental law describing the evolution of quantum fields, the Schrödinger equation, was probabilistic and therefore not exact!

In short, it is clear that if meteorology aspires to be both a rigorous science and one where users can rely on our outputs (literally, as though their lives depended on them), then we must develop systems that can predict the flow-dependent consequences of inevitable uncertainties in our forecast systems. This is what ensemble prediction and the ECMWF EPS in particular, attempts to do.

### 3. The Met Office Long-Range Forecast Ensemble

My personal involvement in the development of ensemble prediction systems predates work begun at ECMWF on medium-range forecasting. In 1982, I returned to the Met Office from a spell at the University of Washington where I had been working on stratospheric dynamics, and found myself posted to the so-called Synoptic Climatology Branch; synoptic climatology being a field about which I knew almost nothing. However, the branch had had some illustrious past members, notably the renowned climatologist Hubert Lamb who went on to found the Climatic Research Unit at the University of East Anglia, and this gave me some confidence that there was worthwhile science to be done. One of the practical duties of the Synoptic Climatology Branch was to produce 30-day forecasts for 10 districts of the United Kingdom, for a number of external fee-paying customers, such as utility companies. Such forecasts were mainly made using statistical-empirical models (Folland and Woodcock, 1986) whose input were current circulation patterns, and, importantly, sea surface temperatures from regions around the world. The output from such models was probabilistic in character. For example, one type of output was the probability, over the coming month, of Lamb Weather Types – what would now be referred to as circulation regimes.

Over the years, the Synoptic Climatology Branch had become somewhat detached from the work of other forecast branches at the Met Office, whose work from the 1960s onwards had been revolutionised by the development of numerical models of the atmosphere. The task of me and my team was to investigate whether numerical models might also play a role in long-range forecasting, from monthly to seasonal timescales. On the seasonal timescales this involved studying the atmospheric model response to El Nino and related sea surface temperature anomalies (Palmer and Mansfield, 1986a, b). However, two important papers suggested that numerical forecasting on the monthly timescale was a worthwhile pursuit in its own right: Shukla (1981) had suggested that there was some potential predictability in spatially and temporally averaged fields on the monthly timescale, and Miyakoda (1983) had indicated that the severe US winter of 1976/77 was predictable on the monthly timescale using the latest generation of weather/climate models.

Work had already started in this direction using the Met Office hemispheric 5-level model (Mansfield, 1986). However, the arrival of the new global 11-level model (and a new supercomputer) presented new opportunities. The early form of this model had a profound westerly bias that made it quite useless for long-range forecasting and provoked the development of a parametrisation of a hitherto unparametrised sub-grid process - orographic gravity waves (Palmer et al, 1986). I will return to this parametrisation below in the discussion about stochastic parametrisation.

Of course, it made no sense to base a monthly forecast on a single deterministic integration of the 11-level model. Although this was implied by Lorenz's 1963 model, it became a familiar fact of the matter to those of us working with complex numerical models, that pairs of integrations started, say 12 hours apart, would diverge radically on monthly timescales. At the very least, one should perform multiple runs from consecutive initial conditions and somehow synthesise the results. How should we do this? One approach had been suggested by Shukla: look at only the large-scale low-frequency modes of variability by applying spatial and temporal averaging to the fields. Another approach had been suggested by Dalcher et al (1987): simply average together the integrations from consecutive analyses. The technique was called lagged-average forecasting, the idea being that unpredictable scales would be filtered out by this averaging process. Both spatial, temporal and lagged averaging would tend to reduce the root-mean-square error of the forecast. However, there is a penalty to be paid by such averaging: extreme events with relatively poor predictability will necessarily be filtered out of the forecast. And yet it is precisely these events that the user needs to be warned about!

In the case of the monthly ensembles in the Synoptic Climatology Branch, there was an obvious alternative approach: since the empirical model output was probabilistic in nature, and since our goal was to try to combine the empirical and numerical model output, we should synthesise the ensemble forecast output probabilistically. Hence running 7 forecasts made from consecutive 12-hour analyses, we created a rudimentary ensemble-based probability forecast based on a cluster analysis of the 7 circulation patterns that were predicted. Although the forecast from the earliest initial condition was likely to be the least skilful *a priori*, the actual differences in skill between the 7 ensemble members, on these monthly timescales, was too small to be concerned about.

And so, starting in November 1985, the Met Office quasi-operational probabilistic ensemble forecast system was born (Murphy and Palmer, 1986) and started to complement the statistical methods (Folland and Woodcock, 1986). I believe this was the world's first real-time probabilistic ensemble forecast system, based on numerical models of the atmosphere.

**4. The ECMWF Ensemble Prediction System**

The moments in one's life when one suddenly grasps some scientific truth, the so-called Eureka moments, are rare indeed. For me, the time when I thought, sometime in the mid 1980s - Why aren't we designing probabilistic ensemble forecast systems for the short and medium range as well as the extended range? – was one such. In hindsight, and in the light of Fig 1, it is an obvious idea and hardly up there with the great scientific discoveries of the 20th Century. However, in the 1980s, the demarcation line between short and medium-range predictions within the deterministic limit, and monthly and seasonal predictions beyond it, was very strong indeed.

The idea that medium-range forecast uncertainty had to be somehow predicted had been recognised by the Dutch meteorologist and fluid dynamicist Henk Tennekes who famously said that "no forecast is complete without a forecast of the forecast skill" (Tennekes et al, 1987). As a result, much of the thinking in the second half of the 1980s, by those concerned with medium-range predictability, was focussed on trying to devise some way of predicting whether the latest deterministic forecast was going to be skilful or unskilful. If this could be achieved it would of course be valuable. However, it seemed to me that the value of "forecasting the forecast skill" would always be rather limited. The famous October 1987 storm provides a good case in point. Merely knowing that a short-range deterministic forecast of a relatively benign day was likely to have a large rms error, would itself not provide a warning of the possibility of an unprecedented storm with hurricane force gusts in the south of England. This is where a reliable ensemble would have provided much more valuable information than a mere forecast of deterministic forecast skill.

*a) Singular vectors*

These conceptual issues notwithstanding, a key practical problem in designing a medium-range ensemble is how to produce a viable set of initial perturbations which would adequately sample initial uncertainty. The time-lagged approach based on consecutive analyses was clearly going to be of very limited value if one sought ensemble sizes of 30 or more (needed to estimate probabilities without large sampling errors, especially of extreme events), since then one would have to use analyses that were so out of date as to be quite useless. One would like perturbations consistent with analysis error - but how to construct these in practice? Clearly these perturbations would be small in scale, since it is on scales smaller than the resolution of the observation network that the analysis will be most uncertain – compounded by the fact that model error is likely also to be dominant on such small scales. And yet the unrealistically diffusive dynamics of models on small scales will damp such small-scale perturbations. Tony Hollingsworth, Head of Research at ECMWF for many of the years I was at ECMWF, had discovered this when he ran forecast twin experiments, adding spatially and temporally uncorrelated noise to the initial conditions. Instead of the noise growing as one would have expected from Lorenz's butterfly effect, the model's diffusive terms damped the noise exponentially. Such "under-dispersiveness" has dogged ensemble forecasting ever since.

When I began working at ECMWF, I did not know how to solve this problem, and so temporarily shelved it in favour of studying plausible schemes that could "forecast the forecast skill". One such was based on an analysis of patterns of large-scale variability in 500hPa height which were maximally correlated with variations in medium-range deterministic forecast error. Using a linear regression analysis between medium-range RMS forecast error and EOFs of the forecast flow, it was found that the Pacific/North American (PNA) teleconnection pattern was a particularly good predictor of medium-range forecast error over the Pacific and North American sector (Palmer and Tibaldi, 1988) – an interesting result though of limited practical value for European weather forecast users!

In trying to understand the dynamical origin of this result, I adapted Adrian Simmons' barotropic model (Simmons et al, 1983) with basic states which included either positive and negative PNA patterns added to the time-mean climatology. Perturbations, initially upstream of the PNA region, were found to grow much more strongly on the negative PNA basic state, suggesting the forecast error results from the regression analysis could be understood in terms of a large-scale barotropic instability. I began writing a paper on this (Palmer, 1988) but, as a last thought, computed the dominant eigenvalues of linear growth for the positive and negative PNA basic states. It was "obvious" that since the negative PNA basic state was more unstable it would have the larger eigenvalue. The opposite turned out to be the case. For some weeks, I was bamboozled, believing there must be a bug in the code.

At this point, I recalled one of Brian Farrell's seminars some years earlier when he was visiting the Met Office, claiming that normal-mode instability was largely irrelevant for studying transient growth of baroclinic waves in midlatitudes. Brian described the "Orr Mechanism" as applied to the Charney baroclinic instability problem (Farrell, 1989). From a mathematical point of view, the Orr mechanism describes transient growth which is not bounded above by eigenvalue growth, and arises because the linear evolution equations for the Charney problem are not of Sturm-Liouville type and the corresponding linear dynamical operators are not self-adjoint (essentially because of shear in the basic state). That is to say, transient growth can occur if the eigenvectors of the linear operator are not orthogonal, even if the leading eigenvalue is negative. Fig 2 shows a schematic illustration of this process. Thinking about this I realised that this mechanism was endemic to all linear dynamical problems where the background state has shear (and hence where the background state cannot be commuted with the gradient operator in the linearised form of the nonlinear advection term). In some sense such non-normality is therefore a residual manifestation in the linearised problem, of the underpinning nonlinear problem. That is to say, even though they are linear, singular vectors provide a partial manifestation of the nonlinearity of the underpinning system – I will return to this point later. With a PhD student of Brian Hoskins, Zuojun Zhang, we examined the non-normality of the eigenvectors of the positive and negative PNA basic states and found that a measure of this non-normaility (the inverse of the cosine of the angle between an eigenvector and its adjoint) provided an explanation of the results found in the barotropic model integrations.

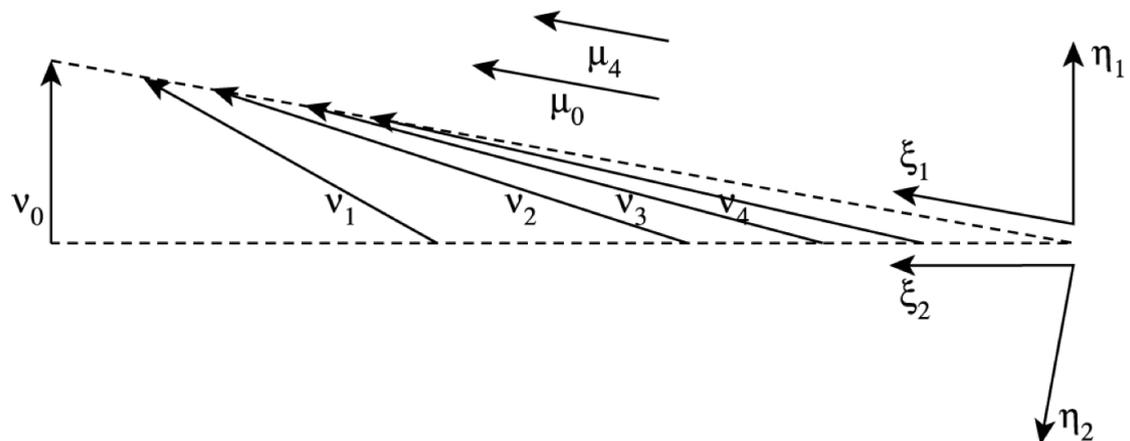

Figure 2. This shows schematically how singular-vector growth can dominate eigenvector growth for non-self-adjoint (i.e. non-Sturm-Liouville) linearised dynamical systems. Here we consider a 2- dimensional linear dynamical system with decaying eigenvectors $\xi_1$, $\xi_2$ and adjoint eigenvectors $\eta_1$, $\eta_2$. We assume $\xi_1$ decays more slowly than $\xi_2$. By construction $\eta_1$ is orthogonal to $\xi_2$ and $\eta_2$ is orthogonal to $\xi_1$. A unit-amplitude initial perturbation $\mu_0$ which projects onto the $\xi_1$ direction will decay as shown. However, a unit-amplitude singular-vector perturbation $\nu_0$ proportional to the difference $\xi_2 - \xi_1$ will grow (at least over finite time) because, by construction, the tip of the arrow will evolve more slowly than the tail. The propensity for such non-modal growth is determined by the extent to which the angle between an eigenmode and its adjoint differs from zero. Here the angles are not far from the maximal value of 90° associated with eigenvector degeneracy.

A more general way to study this type of non-modal growth was through the so-called singular vectors of the linearised dynamics. Singular value decomposition goes back to the work of some of the great mathematicians of the 19th Century such as Eugenio Beltrami and Camille Jordan but (I believe) was first referred to in a meteorological context by Lorenz (1965). If we write the linearised dynamics in the propagator form $\delta x(t_1) = A(t_1,t_0)\delta x(t_0)$, then the singular vectors of $A(t_1,t_0)$ are simply the eigenvectors of $A^T(t_0,t_1)A(t_1,t_0)$ at initial time and the eigenvectors of $A(t_1,t_0)A^T(t_0,t_1)$ at evolved time. It is easy to see that $A^T(t_0,t_1)A(t_1,t_0)$ is self-adjoint, even if $A(t_1,t_0)$ is not. Hence the growth of perturbations defined from linear combinations of singular vectors will be bounded above by the corresponding singular values when, as Fig 2 shows, linear combinations of eigenvectors will not be bounded by the corresponding eigenvalues. Together with Franco Molteni, the leading singular vectors were calculated using matrix algorithms from a 3-level T21 quasi-geostrophic model (Molteni and Palmer, 1993). Using an energy metric (that is to say, where the perturbations have unit energy norm) it was found that the singular vectors propagated up-spectrum from sub-synoptic to synoptic scales between initial and evolved time (taken as 2-days). Broadly speaking, this upscale growth process illustrates the Orr mechanism first found by Brian Farrell for baroclinically unstable flow.

This was exactly the type of process needed to offset the erroneous damping effects of small-scale model diffusion. The initial perturbations were small scale, as required by data assimilation, and they were guaranteed to evolve into meteorologically balanced perturbations. That is to say, these singular vectors could indeed provide a viable means to represent initial perturbations for a medium-range ensemble forecast system. Initial tests (Mureau et al, 1993) gave good results. The singular vectors were then calculated from the primitive equation forecast model (Buizza et al, 1993; Buizza and Palmer, 1995) using the adjoint model developed for 4DVAR (and an iterative Lanczos scheme to compute the leading singular vectors) and this became the basis of the ECMWF Ensemble Prediction System which went operational in late 1992 (Palmer et al, 1992; Molteni et al, 1996).

At the same time, NMC/NCEP implemented an operational medium-range ensemble forecast system but based on so-called breeding vectors, rather than singular vectors (Toth and Kalnay, 1993, 1997). A breeding vector can be likened to that of finding an eigenvector through the power method: take a random perturbation, grow it, rescale it and repeat indefinitely. In the following years, a number of debates ensued with Eugenia Kalnay and Zoltan Toth about the merits of singular versus breeding vectors. One argument sometimes used to support breeding over singular vectors was that breeding vectors are metric independent whilst singular vectors depend strongly on the choice of metric (particularly at initial time). However, I have always viewed this dependency as a strength, not a weakness, of the singular-vector method. In particular, there is a natural (initial-time) metric to use if the singular vectors are, as they are here, designed to probe the predictability of weather: it is the analysis-error metric and defined by the analysis error covariance matrix. In a number of studies (Gelaro et al, 1998: Palmer et al, 1998; Barkmeijer et al, 1999), it was shown that the simple energy metric provided a good approximate representation of the more expensive analysis-error covariance matrix provided by variational data assimilation.

A key feature of the fastest-growing energy-metric singular vectors (shared with analysis-error singular vectors) but not with, say, enstrophy-metric singular vectors, was that the initial vectors were typically sub-synoptic, whilst the evolved singular vectors were synoptic scale and larger. That is to say, the singular vectors captured the essential non-modal upscale growth of analysis-to-forecast error from sub-synoptic scale, where uncertainty was largest, to synoptic and larger scales where most of the energy in the atmosphere resides. This can be contrasted with the breeding vector methodology, where, as mentioned, the perturbations are simply renormalized at every cycle time. Such renormalisation does not mimic the way in which observations transform a first-guess error at the end of one data assimilation cycle to an analysis error at the end of the next data assimilation cycle. In particular, such renormalisation does not capture the way observations strongly damp first-guess errors on scales larger than the observation network, but leave unchanged first-guess errors on scales much smaller than the observation network. In other words, whilst data assimilation can be expected to rotate the error field in wavenumber space, this is not to be expected from a breeding cycle. This was my main argument why singular vectors were, theoretically at least, preferable to breeding vectors.

On the other hand, computing singular vectors in anything other than a simple model context (i.e. whether the number of degrees of freedom was sufficiently small to allow explicit matrix representations) requires an adjoint model. At ECMWF this was possible because of the development of 4DVAR. At other operational centres this was not possible.

Of course the proof of the pudding is in the eating. Does it actually make any difference? Fig 3 compares the spread and skill of the current ECMWF EPS with those from the UK Met Office and NCEP (neither of whom use singular vectors). It can be seen that the latter are notably under-dispersive. This is also true in the Southern Hemisphere (not shown).

This result does not prove that singular vectors are playing a decisive role. After all, a component of the EPS initial perturbations nowadays arises from the use of an ensemble of 4DVAR data assimilations. However, Fig 4 shows that singular vectors are still a critical component of the EPS. Without these singular vectors the ensemble would also be strongly underdispersive and more similar to the other operational ensemble systems shown in Fig 3. A possible reason for the continued need for singular vectors is discussed in Section 5.

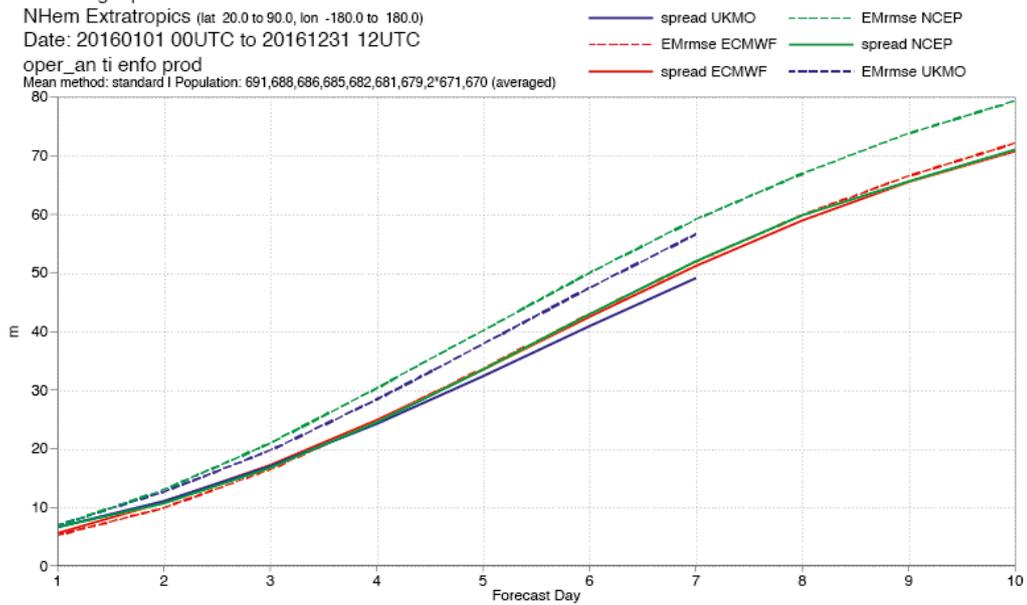

*Figure 3. A comparison of three operational ensemble prediction systems for NH 500hPa geopotential height. Solid lines show ensemble spread and dashed lines show ensemble-mean RMS error. The ECMWF ensemble is shown in red, the NCEP ensemble is shown in green and the UK Met Office ensemble is shown in blue. The spread and skill of the ECMWF ensemble are extremely well balanced. However, the other two operational ensembles are underdispersive (with error exceeding spread). Martin Janousek, personal communication, 2017.*

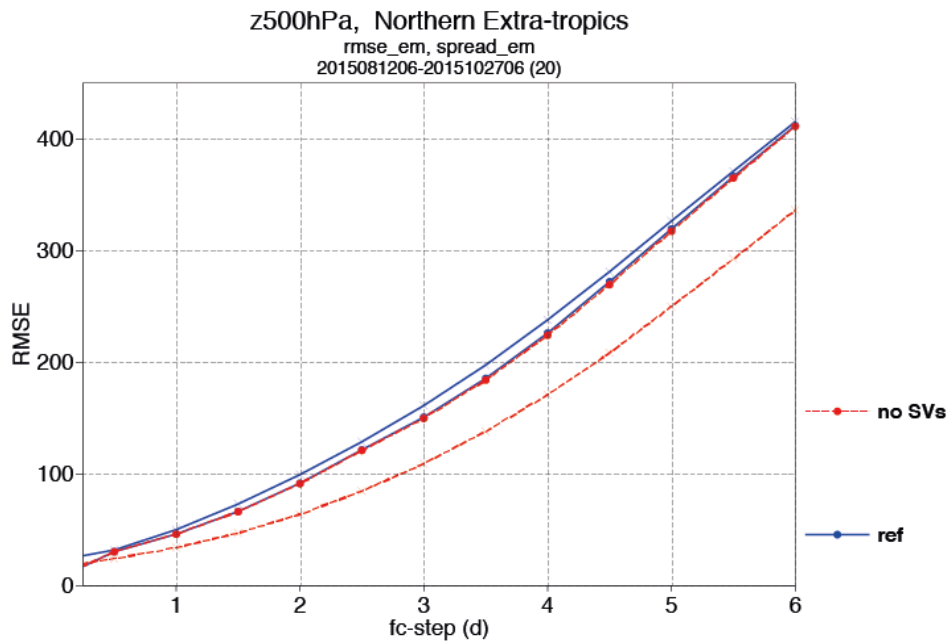

*Figure 4. Estimates of spread and skill for Z500 NH extratropics from the current ECMWF ensemble, with and without singular vectors. Blue shows results with singular vectors, red without. Despite the use of ensemble data assimilation in generating initial perturbations, singular vectors remain a vital element in accounting for the reliability of the current ECMWF ensemble prediction system. This result illustrates the fact that ensembles of 4DVAR data assimilations do not themselves produce adequate spread. Simon Lang, personal communication, 2017.*

b) Stochastic physics

It was clear from the very earliest stage that initial perturbations were not going to be sufficient to create a reliable ensemble. By day 7 the ensemble had become somewhat under-dispersive in the extratropics, whilst by day 3 it had become very under-dispersive in the tropics. One could inflate the singular vector amplitudes so that the ensemble spread at day 7 was correct. However, the ensemble would then be over-dispersive in the early stages of the forecast. This was not the right way to proceed.

The answer was fairly obvious: there was no representation of model uncertainty in these early ensembles. In the early 1990s, this was something I would discuss with Martin Miller, Head of the Physical Aspects Section and later Model Division at ECMWF (the most decisive discussion occurring when we were both returning from a week's golfing vacation). At the time, the Met Office were pursuing a "perturbed parameter" approach to representing model uncertainty in climate change ensembles. Martin and I had different reasons for believing this was not the right approach for the EPS. For me the key reason arose from my involvement in the development of the orographic gravity wave drag parametrisation designed to make the Met Office 11-layer model a reasonably viable tool for ensemble monthly forecasting (see above). However, in developing this parametrisation a number of very significant and substantial approximations had to be made, not least that the propagation of the waves was based on the simplest type of linear theory. The idea that one could represent the uncertainties in these approximations (e.g. nonlinear effects) by merely perturbing the basic parameters of the parametrisation, keeping the closure formulae fixed, seemed to me quite preposterous (and still does!). That is to say, a scheme which perturbs parameters but does not perturb the closure scheme itself, is likely to produce very conservative (and hence under-dispersive) estimates of uncertainty.

Martin Miller's principal reason for disliking perturbed parameters was somewhat different. For him, representing the sub-grid tendencies correctly in a gridbox, where in reality let's say deep convection was occurring, would not merely require a realistic convective parametrisation, it would also require realistic parametrisations of cloud condensation processes, radiation and boundary layer turbulence. Crucially, these must be in proper balance with respect to one another. Perturbing individual parameters in individual parametrisation schemes, but keeping other parameters fixed, would tend to destroy this balance and hence destroy the realism of the total parametrised tendency.

It was on this basis that the alternative idea of making the total parametrisation tendency stochastic arose. Martin invited me to talk about the notion of stochastic parametrisation in a workshop on Convection Parametrisation (Palmer; 1996, 2001). After this workshop, Martin and I worked together to develop the scheme that is now known as SPPT (Stochastically Perturbed Parametrisation Tendencies), adding a new stochastic term to the prognostic equations, whose amplitude is proportional to the total parametrised tendency for the variable in question. This was tested in the EPS and results were positive (Buizza et al, 1999). Coarse-grain studies (Shutts and Palmer, 2006) subsequently lent some support to such "multiplicative-noise" schemes. A more refined scheme was developed (Palmer et al, 2008) and it was shown that SPPT improved probabilistic scores in the tropics by about 4 days. This shows the power of relaxing a key incorrect assumption present since the earliest days of numerical weather prediction: that just because the underlying equations of motion are deterministic, the numerical representation of these equations must also be deterministic.

By perturbing the whole tendency, SPPT respects the holism that Martin Miller himself felt was so vital to represent when doing uncertainty quantification. Of course, one should treat this as a zeroth order approximation and there are undoubtedly circumstances where some partial or even full decoupling does make sense. However, occasionally it is commented that by perturbing the whole tendency, SPPT is somehow *ad hoc* and therefore not physically based. Nothing could be further from the truth; just because a scheme is simple, it is not necessarily simplistic! The fact that it has so far not been possible to improve skill scores with alternative model error schemes suggests that SPPT is doing something right.

**5. From Hours to Decades. Job done?**

As discussed, the origins of ensemble forecasting arose in extended range prediction and from there, it percolated into medium-range forecasting. On longer timescales, it has been used to develop probabilistic seasonal and decadal forecast systems based on global coupled models and on shorter timescales has been used to develop probabilistic short-range forecast systems using very high-resolution limited-area models. Reviews of the use of ensemble methods across the timescales have been given elsewhere (Palmer, 2000: Palmer et al, 2005).

Job done? No! These probabilistic forecasts are only going to be useful for decision making if the forecast probabilities are reliable – that is to say, if forecast probability is well calibrated with observed frequency. Whilst most medium-range forecast ensemble forecast products are reliable, on longer timescales they become less so. Weisheimer and Palmer (2014) documented the fact that on the seasonal timescale, many seasonal-mean precipitation forecasts are far from reliable and all the evidence suggests that this is due to the ensemble systems misrepresenting model error rather than observation error. As such, one must surely be cautious about the reliability of regional climate-change projections of precipitation (Palmer et al, 2008).

Such results argue for increasing model resolution to try to minimise the systematic model errors that develop on the seasonal timescale – on the basis that representing processes using the proper laws of physics than by highly uncertain parametrisations is surely a good thing. Indeed, there are very good reasons for aspiring to produce a 1km global ensemble forecast system (Palmer, 2016): not least two of the key physical processes in the atmosphere (deep convection and orographic gravity-wave drag) become resolved.  I strongly believe that developing the 1km model will be essential if we are to eliminate the model biases that have plagued our field from the very start.

From a computational point of view, ensembles were an easy sell since they are "embarrassingly parallel" and therefore well suited to massively parallel supercomputers. Increasing resolution is much more problematic, given the projected end of Moore's Law. However, here ensemble forecasting, and stochastic parametrisation in particular, has suggested a new way forward. For example, if parametrisations are fundamentally stochastic in nature, then why are we representing all the variables in a weather or climate model with 64 bits? Is there real information in all these bits, or can some be removed without loss of accuracy. If so how many? If substantial, the computer resources saved can be redeployed to increase model resolution. In turns out that in terms of electrical power,  the data transport costs of 64-bit representations are large and wasteful and ensemble skill is unchanged moving (almost entirely) to single precision (Düben et al, 2014; Váňa et al, 2017). There are strong indications we can go much further, with large parts of the model coded using 16-bit half-precision floating point reals. Here we may be able to take advantage of a new class of mixed-precision processor, designed primarily with the machine learning community in mind.

One hint as to whether higher-resolution models will produce more reliable forecasts can be obtained by looking at the short-range ensemble forecasts using high resolution limited-area models. Unfortunately, here results are not encouraging. Fig 5 shows a reliability diagram for radar reflectivity (indicating convective precipitation) from the Community Leveraged Unified Ensemble (CLUE): Jirek et al (2016). CLUE is a 20-member multi-model ensemble of limited-area models with 12 km horizontal resolution. As Jirek et al comment, and as can be seen from Fig 5: "overall, the ensembles were very underdispersive for reflectivity forecasts".

At the very least, it would seem that these high resolution short-range ensembles could benefit from singular vector perturbations just as the medium-range predictions have. However, this raises an important question. Why exactly are singular vectors so important? After all, model resolution has increased to the point where we are not seeking to add perturbations directly to the model's highly diffusive scales. In this respect, it is worth commenting on an interesting new result from Wayne and Durran (2018) who studied error growth in a mesoscale model with variable resolution. Initial perturbations with a fixed scale of around 100km were found to grow faster (and propagate to larger scales faster) the smaller was the resolution of the model. This itself may not appear too surprising, but the effect was still found when the resolution of the model was increased from 2km to 1km – two orders of magnitude smaller in scale than the perturbation itself. The authors conclude this is a consequence of nonlinearity in an atmosphere with a -5/3 spectrum (Lorenz, 1969): that perturbations propagate nonlinearly downscale and then back up to the mesoscale, reflecting the butterfly effect as Lorenz originally intended it to mean (Palmer et al, 2014). If this explanation is correct, it provides some justification for the comment above, that the non-modal nature of singular vectors describes in a linearised framework a process whose ultimate explanation lies in nonlinear advection in a sheared flow and in the "real" butterfly effect (Palmer 2014). This process and its implications for understanding the impact of model resolution on large-scale error certainly require further attention.

My point in highlighting this issue, is that it should be a concern not just to short-range weather forecasters, but also to climate modellers seeking to develop high resolution models that can potentially simulate convective systems better. For example, the CLUE ensembles lie in the "grey zone" where convective systems are neither resolved nor parametrised adequately. Do these reliability diagrams suggest one should avoid the grey zone as a target resolution for next generation global models and wait until we can run with a 1km grid? These are important issues for both weather and climate prediction, and highlight a need for scientists from both sides to work closely together.

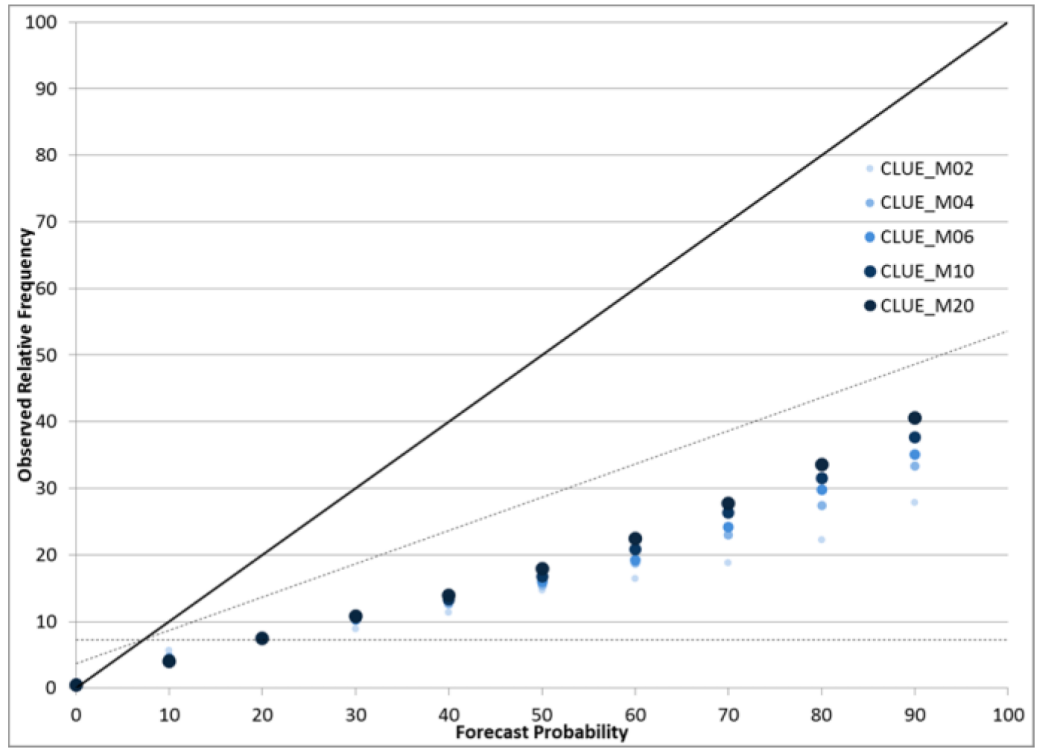

*Figure 5. Reliability diagram for probabilistic reflectivity forecasts ≥40 dBZ from the CLUE ensemble size experiment during SFE2016 (From Jirak et al, 2016).*

Before concluding this Section, I would like to make a suggestion for further improving initial ensemble perturbations. Since singular vectors seem unlikely to be removed in the coming years – something I would not have predicted a few years ago -  perhaps we should start to refine the initial singular vector metric and computation little more. In particular, the regional amplitude of SPPT perturbations in the ensembles of data assimilations could itself be used to constrain the initial perturbations, increasing their amplitude in regions where SPPT is especially active, damping in other regions. This reflects the fact that initial error is not just due to observation error; it is also due to model error. Indeed, it many parts of the world (e.g. where mesoscale convective activity is occurring), model error may be the dominant contribution to initial error. Secondly, there is also good reason to include moist diabatic processes into the singular vector computation in the extratropics.  Moist singular vectors are used for perturbations targeted on tropical cyclones (Puri et al, 2001). However, they have never replaced the dry singular vectors used in the extratropics.

## 6. The next 25 Years: Meeting the Sendai Framework Development Goals

Where would I like to see ensemble prediction 25 years from now? Firstly, I hope that long before then, operational weather forecast centres will drop their `high-resolution deterministic' forecast, run in parallel with the ensemble, and focus on a high-resolution ensemble forecast (Palmer, 2012).As discussed, because of the intermittent butterfly effect, deterministic forecasts are by their nature unreliable. As such, the added detail provided by a single high-resolution deterministic run, which cannot by definition be supported by the corresponding ensemble, must be unreliable and therefore not useful for users. (Aspects of high resolution which are slave to lower resolution elements of the flow can be produced by statistical downscaling if necessary). Rather, the route to higher resolution forecasts is through limited area-model ensembles, embedded into the global ensemble, and this is a job for the national met services. It is worth noting in this respect that ECMWF's strategic goal is to develop a 5km ensemble by about 2023, at which point the resolution of the deterministic forecast will be the same as the ensemble.

Indeed, let's assume that in 25 years we will be running global ensembles at 1km resolution and that the sort of problems suggested in Fig 5 have been solved. With such ensembles, we will finally be able to provide reliable estimates of weather risk for severe events (including wind and precipitation intensity), on forecast timescales of a week or so. Why is this so important? One of the goals of the UNISDR Sendai Framework for Disaster Risk Reduction (https://www.unisdr.org/we/coordinate/sendai-framework) is a substantial reduction in mortality due to natural disasters. According to the UNSIDR, over 90% of natural disasters are weather related. In this context, it is a great source of frustration to many of who have worked to improve weather forecasts to see populations suffering a week after some extreme weather event has devastated their neighbourhoods. An example is Typhoon Haiyan which hit the Philippines in 2013. According to the BBC web site (http://www.bbc.co.uk/news/av/world-asia-24958868/help-message-seen-by-a-us-surveillance-aircraft), only a week after the typhoon hit the Philippines, did food and supplies begin to reach survivors. In the days following the event, aid agencies stressed that the logistics of distribution were enormous, and simple messages of "Help! We need food and water" could be seen by surveillance aircraft.

And yet the track of Haiyan was relatively well known a week or so before the typhoon hit land: the ensemble spread was relatively small. This raises the question of whether disaster relief agencies can start to be more proactive when these high-impact events are forecast to hit; to get food, water, medicine and shelter in place before the event hits. The problem, of course is that it is easy to be wise after the event. The critical question is whether weather forecasts are reliable enough to allow disaster relief agencies to make decisions on when to be proactive – starting to act let's say a week ahead of the event - and when the uncertainty is simply too large to do anything but be reactive. Such agencies have limited resources and if they act proactively too often, in cases where the supposed life-threatening does not materialise, then they will go bust.

We can frame the decision process in a schematic and idealistic way. Let $C$ denote the cost of some proposed disaster relief action, without which some loss $L$ would occur. (This of course, suggests one can put a price on human suffering and loss of life, a distasteful but necessary fact of the matter.). Let $p$ denote the forecast probability of the life-threatening event. Then a simple decision theoretic analysis suggests that proactive action should be taken if $p>C/L$ (Murphy, 1969). In practice, it would make sense to add an offset to ensure the disaster relief agency does not go bust after a run of occasions when action is taken but the event does not occur. By definition, this type of objective decision-theoretic analysis can only be performed if reliable probabilities are available and is why developing fully reliable ensemble forecast systems is so crucially important.

Of course, it is not just disaster relief agencies who can benefit from reliable probability forecasts. In a ground-breaking study, Webster et al (2010) showed that farmers in rural Bangladesh readily appreciate the value of probability forecasts for decisions on whether to move their livestock (often their principal assets) to higher ground. In Webster et al (2010) it was shown quite conclusively that these farmers understood well the concept of probability, scotching the myth that probability is too complex a concept for the average person to understand.

In summary, in 25 years, my expectation is that 1km global ensemble forecasts will be used objectively for a range of applications and where the probabilities will be intermediate from the weather forecast model to the application model (Palmer, 2002). Indeed, going further, I can imagine a situation where private companies, as well as disaster relief agencies, embed their impact models into the ensemble forecast system as it runs, enabling probabilities of direct interest to the users to be produced without having to output petabytes of high-resolution high-frequency meteorological data. (The private companies would pay to have their application models embedded into the forecast models in this way, and would pay for extra computational capacity to ensure their application model does not slow down operations. Non-disclosure agreements would ensure that the impact-model code remains confidential.). The user-relevant probabilities would then be sent to the users to enable a definite non-probabilistic course of action.

## 7. Conclusions

In this paper, a personal account has been given of the background to the development of the ECMWF ensemble prediction system. Although it is now over 25 years since the EPS became operational, it was far from clear, for many years, whether it would ever see its 25th birthday. Referring to the EPS, my own boss used to introduce meetings with ECMWF Member State representatives saying that their forecasters, for whom this represented a major change in approach, should "use it or lose it". For a number of years I worried that the whole thing would be wound up as an exercise that, whilst worthy, was not felt to add sufficient value to the practice of weather forecasting. It was only relatively recently, at the WMO/WWRP/THORPEX World Weather Open Science Conference in Montreal in 2014, when I finally realised that ensemble forecasting had finally become an irreversible part of meteorology: talk after talk from both forecasters and researchers alike used ensemble methods as they were second nature. It reminded me of the old adage that you never convince your sceptical colleagues, it's just that a new generation comes along for whom there was never any real alternative.

Over the next 25 years I expect that the current practice of running both a high-resolution deterministic forecast and an ensemble at lower resolution, will slowly disappear in favour of high-resolution (ultimately 1km) global ensembles (feeding yet finer-scale regional ensembles). This will finally give us the sort of quantitatively reliable risk assessments that will transform business and societal decision making towards largely objective procedures – e.g. when to target disaster aid relief ahead a projected event, and therefore start gearing up a week or more before the event is projected to hit. This transformation is an essential tool in making society worldwide become more resilient to extremes of weather - such extremes only becoming more commonplace as human-induced climate change takes hold. In this way, policy makers will recognise the value of high resolution weather and climate models, and fund their development commensurate with their importance. As a result, standing on the shoulders of the meteorological giants of the past, I expect meteorology to enter a golden age as has never been seen before.

## Acknowledgements

As mentioned, the development of the ECMWF medium-range Ensemble Prediction System goes far beyond the team directly involved in the development of the ensemble. However, it is not feasible to thank everyone in person here. Hence below are the specific scientists who worked with me to develop the ECMWF EPS. Without them, the EPS would not exist in its present form: Jan Barkmeijer, Roberto Buizza, Philippe Chapelet, Dennis Hartmann, Renate Hagedorn, Martin Leutbecher, Jean-Francois Mahfouf, Martin Miller, Franco Molteni, Robert Mureau, Thomas Petroliagis, Kamal Puri, David Richardson, Glenn Shutts, Stefano Tibaldi and Joe Tribbia.

I would also like to acknowledge the contributions of Simon Lang and Martin Janousek in supplying Figs 3 and 4 respectively.

## References

Barkmeijer, J., R. Buizza and T.N. Palmer, 1999: 3D-Var Hessian singular vectors and their use in the ECMWF Ensemble prediction system. Q.J.Roy.Meteorol.Soc., 125, 2333-2351.

Barkmeijer, J., R. Buizza, T.N.Palmer and K.Puri, 2001: Tropical singular vectors computed with linearised diabatic physics. Q.J.R.Meteorol.Soc., 127, 685-708.

Buizza, R., J. Tribbia, F. Molteni and T.N. Palmer, 1993: Computation of optimal unstable structures for a numerical weather prediction model. Tellus, 45A, 388-407.

Buizza, R. and T.N. Palmer, 1995: The singular vector structure of the atmospheric global circulation. J.Atmos.Sci., 52, 1434-1456.

Buizza, R., M.J. Miller and T.N. Palmer, 1999: Stochastic simulation of model uncertainties in the ECMWF Ensemble Prediction System. Q.J.R.Meteorol. Soc., 125, 2887-2908.

Dalcher, A., E. Kalnay and R.N.Hoffman, 1987: Medium Range Lagged Average Forecasts. Mon. Wea. Rev., 116, 402-416.

Düben, P. D., McNamara, H., & Palmer, T. N., 2014: The use of imprecise processing to improve accuracy in weather & climate prediction. *Journal of Computational Physics*, *271*, 2-18. doi:10.1016/j.jcp.2013.10.042

Ehrendorfer, M.,2006: The Liouville Equation and Atmospheric Predictability. In Predictability of Weather and Climate. Edited by Palmer, T.N. and R. Hagedorn. Cambridge Unversity Press.

Epstein, E.S., 1969: Stochastic-dynamic prediction. Tellus, 21, 739-759.

Farrell, B.F., 1989: Optimal excitation of baroclinic waves. J. Atmos. Sci., 46, 1193-1206.

Folland, C.K and A. Woodock, 1986: Experimental monthly long-range forecasts for the United Kingdom. Part 1. Description of the forecast system. The Meteorological Magazine, 115, 301-318.

Gelaro, R., R. Buizza, T.N. Palmer, E. Klinker, 1998: Sensitivity analysis of forecast errors and the construction of optimal perturbations using singular vectors. J.Atmos.Sci.,55, 1012-1037.

Jirak, I.L., A. J. Clark, C. J. Melick, S.J.Weiss, 2016: Investigation of the Impact of Convection-Allowing Ensemble Size for Severe Weather Forecasting. Proceedings of the 28th Conference of the American Meteorological Society on Severe Local Storms.

Leith, C.E., 1965: Numerical simulations of the Earth's atmosphere. *Methods in Computational Physics*, eds. B. Alder, S. Fernbach, and M. Rotenberg (New York: Academic Press), 1-28.

Lorenz, E.N., 1963: Deterministic Non-Periodic Flow. J. Atmos. Sci., 20, 130-141.

Lorenz, E.N., 1965: A study of the predictability of a 28-variable atmosphere model. Tellus, 17, 321-333.

Lorenz, E.N., 1969: The predictability of a flow which possesses many scales of motion. Tellus, 21, 289-307.

Mansfield, D. 1986: The skill of dynamical long-range forecasts, including the effect of sea-surface temperature anomalies. Q. J. Roy. Meteorol. Soc., 112, 1145-1176.

Mintz, Y., 1965: Very long-term global integrations of the primitive equations of atmospheric motion. Proc. WMO/IUGG Symposium on Research and Development Aspects of Long-Range Forecasting. WMO Tech. Note No 66, 141-167

Miyakoda, K., T. Gordon, R. Caverly, W. Stern, J. Siritus and W. Bourke, 1983: Simulation of a blocking event in January 1977. Monthly Weather Review, 111, 846-869.

Molteni, F. and T.N. Palmer, 1993:  Predictability and finite-time instability of the northern winter circulation.  Q.J.R.Met.Soc., 119, 269-298.

Mureau, R., F. Molteni and T.N. Palmer, 1993: Ensemble prediction using dynamically-conditioned perturbations. Q.J.R.Met.Soc., 119, 299-323.

Molteni, F., R. Buizza, T.N. Palmer, and T.Petroliagis, 1996: The ECMWF ensemble prediction system: methodology and validation. Q.J.R.Met.Soc., 122, 73-119.

Murphy, A. H., 1969: On expected-utility measures in cost-loss ratio decision situations. Journal of Applied Meteorology, 8, 989-991.

Murphy, J. and T.N. Palmer, 1986: A real time extended-range forecast by an ensemble of numerical integrations. Met.Mag. 115, 337-348.

Palmer, T.N., and D.A. Mansfield, 1986: A study of wintertime circulation anomalies during past El Niño events, using a high resolution general circulation model. II. Variablility of the seasonal mean response. Q.J.R.Met.Soc., 112, 639-660.

Palmer, T.N., and D.A. Mansfield, 1986: A study of wintertime circulation anomalies during past El Niño events, using a high resolution general circulation model. I. Influence of Model Climatology, Q.J.R.Met.Soc., 112, 613-638.

Palmer, T.N., G.J. Shutts and R. Swinbank, 1986: Alleviation of a systematic westerly bias in general circulation and numerical weather prediction models through an orographic gravity wave drag parameterization. Q.J.R.Met.Soc., 112, 1001-1031.

Palmer, T.N. and S. Tibaldi, 1988: On the prediction of forecast skill. Mon.Wea.Rev., 116, 2453-2480

Palmer, T.N., 1988: Medium and extended range predictability, and stability of the PNA mode. Q.J.R.Met.Soc., 114, 691-713.

Palmer, T.N. F. Molteni, R. Mureau, R. Buizza, P. Chapelet, J. Tribbia, 1992: Ensemble Prediction. ECMWF Technical Memorandum No 188.

Palmer. T.N., 1996. Thoughts on parametrizing scales that are only somewhat smaller than the smallest resolved scales. ECMWF workshop on: "New insights and approaches to convective parametrisation."

Palmer, T.N., R.Gelaro, J. Barkmeijer and R. Buizza, 1998: Singular vectors, metrics, and adaptive observations. J. Atmos.Sci., 55, 633-653.

Palmer. T.N., R. Buizza, F. Doblas-Reyes, T. Jung, M. Leutbecher, G.J. Shutts, M. Steinheimer, A. Weisheimer, 2008 Stochastic parametrisation and model uncertainty. ECMWF Technical Memorandum No. 598.

Palmer, T.N., 2000: The prediction of uncertainty in weather and climate forecasting. Rep. Prog. Phys., 63, 71-116.

Palmer, T.N., 2001: A nonlinear dynamical perspective on model error: a proposal for nonlocal stochastic-dynamic parametrisation in weather and climate prediction models. Q.J.R.Meteorol.Soc., 127, 279-304.

Palmer, T.N., 2002: The Economic value of ensemble forecasts as a tool for risk assessment: from days to decades (The Royal Meteorlogical Society 2001 Symons Memorial Lecture). Q.J.R.Meteorol.Soc. , 128, 747-774.

Palmer, T.N., G.J.Shutts, R. Hagedorn, F.J. Doblas-Reyes, T. Jung and M. Leutbecher, 2005: Representing Uncertainty in Weather and Climate Prediction. Annual Reviews of Earth and Planetary Sciences, 33, 163-194.

Palmer, T.N., F.-J. Doblas-Reyes, M. Rodwell, A. Weisheimer, 2008. Towards Seamless Prediction: Calibration of Climate Change Projections Using Seasonal Forecasts. Bull. Am Met. Soc., 89, 459-470.

Palmer, T.N., 2012: Towards the probabilistic Earth-system simulator: a vision for the future of climate and weather prediction, Quart. J. Roy. Meteor. Soc., 138, 841-861.

Palmer, T. N., Doering, A., & Seregin, G. , 2014: The real butterfly effect. *NONLINEARITY*, *27*(9), R123-R141. doi:10.1088/0951-7715/27/9/R123

Palmer, T. N. ,2016: A personal perspective on modelling the climate system. *Proceedings. Mathematical, physical, and engineering sciences / the Royal Society*, *472*(2188), 20150772.

Palmer, T., 2017: The primacy of doubt: Evolution of numerical weather prediction from determinism to probability, J. Adv. Model. Earth Syst.,9, 730–734, doi:10.1002/2017MS000999.

Puri, K., J. Barkmeijer and T.N.Palmer, 2001: Ensemble prediction of tropical cyclones using targeted diabatic singular vectors. Q.J.R.Meteorol.Soc., 127, 709-731.

Shukla, J. 1981: Dynamical predictability of monthly means. J. Atmos.Sci., 38, 2547-2572.

Shutts G. J. and T.N. Palmer, 2006: Convective forcing fluctuations in a cloud-resolving model: relevance to the stochastic parametrization problem. J. Clim., 20, 187-202

Simmons, A.J., Wallace, J. M. and G.W. Branstator, 1983: Barotropic wave propagation and instability and atmospheric teleconnection patterns. J. Atmos. Sci., 40, 1363-1392.

Smagorinsky, J., 1963: General circulation experiments with the primitive equations. Mon. Wea. Rev., 92, 99-164.

Tennekes, H., Baede, A. P. M. and Opsteegh, J. D., 1987: Forecasting forecast skill. In: Proceedings ECMWF Workshop on Predictability, Reading, April 1986

Toth Z. and Kalnay, E., 1997: Ensemble Forecasting at NCEP and the Breeding Method. Mon. Wea. Rev., 125, 3297-3319.

Toth, Z., and Kalnay, E., 1993: Ensemble forecasting at NMC: The generation of perturbations. *Bull. Amer. Meteor. Soc.,* **74,** 2317– 2330.

Váňa, F., Düben, P., Lang, S., Palmer, T., Leutbecher, M., Salmond, D., & Carver, G. (2017). Single precision in weather forecasting models: An evaluation with the IFS. *Monthly Weather Review*, *145*(2), 495-502. doi:10.1175/MWR-D-16-0228.1

Wayne, J.A. and D.R.Durran, 2018: Do higher resolution models produce more rapid growth of initial-condition errors in simulations of deep convection. J.Atmos.Sci., To appear.

Webster, P.J., Jun Jian, Thomas M. Hopson, Carlos D. Hoyos, Paula A. Agudelo, Hai-Ru Chang, Judith A. Curry, Robert L. Grossman, Timothy N. Palmer, A. R. Subbiah, 2010: Extended-Range Probabilistic Forecasts of Ganges and Brahmaputra Floods in Bangladesh. Bull. Amer. Meteor. Soc., 91, 1493–1514, doi: 10.1175/2010BAMS2911.1.

Weisheimer, A., & Palmer, T. N. (2014). On the reliability of seasonal climate forecasts. *Journal of the Royal Society, Interface / the Royal Society*, *11*(96), 20131162. doi:10.1098/rsif.2013.1162